# Group-velocity-controlled and gate-tunable directional excitation of polaritons in graphene-boron nitride heterostructures

Yuyu Jiang[1,⊥], Xiao Lin[2,⊥,*], Tony Low[3], Baile Zhang[2,4,*] and Hongsheng Chen[1,*]

[1]*State Key Laboratory of Modern Optical Instrumentation, The Electromagnetics Academy at Zhejiang University, Zhejiang University, Hangzhou 310027, China.*
[2]*Division of Physics and Applied Physics, School of Physical and Mathematical Sciences, Nanyang Technological University, Singapore 637371, Singapore.*
[3]*Department of Electrical and Computer Engineering, University of Minnesota, Minneapolis, Minnesota 55455, USA.*
[4]*Centre for Disruptive Photonic Technologies, Nanyang Technological University, Singapore 637371, Singapore.*
[⊥]*These two authors contribute equally to this work.*
[*]*Corresponding author. E-mail: xiaolinbnwj@ntu.edu.sg (X. Lin); blzhang@ntu.edu.sg (B. Zhang); hansomchen@zju.edu.cn (H. Chen)*

**Abstract:** A fundamental building block in nano-photonics is the ability to directionally excite highly squeezed optical mode dynamically, particularly with an electrical bias. Such capabilities would enable the active manipulation of light propagation for information processing and transfer. However, when the optical source is built-in, it remains challenging to steer the excitation directionality in a flexible way. Here, we reveal a novel mechanism for tunable directional excitation of highly squeezed polaritons in graphene-hexagonal boron nitride (*h*BN) heterostructures. The effect relies on controlling the sign of the group velocity of the coupled plasmon-phonon polaritons, which can be flipped by simply tuning the chemical potential of graphene (through electrostatic gating) in the heterostructures. Graphene-*h*BN heterostructure thus present a promising platform toward nano-photonic circuits and nano-devices with electrically reconfigurable functionalities.

# 1. Introduction

Polaritons with high spatial confinement[1-2], such as surface plasmon polaritons in graphene[3-7], and phonon polaritons in hexagonal boron nitride ($h$BN)[8-14], have been intensively explored recent years. Their strong electromagnetic confinement, in conjunction with high quality factor, and electrical tunability, offers interesting opportunities in manipulating light at the extreme nanoscale[15,16]. The high spatial confinement, however, also limits our ability to actively control the excitation direction of these polaritons in a flexible way, which is essential for building the functionalized nano-photonic circuits. While the directional excitation of surface plasmons has been realized via asymmetric structural designs[17-22], their direction of excitation is predefined and cannot be reconfigured on-the-fly if the excitation source is fixed. Until now, the active switching of the excitation direction of polaritons relies primarily on the modulation of the source, including the tuning of the incident angle[23] or the helicity[24-25] (i.e., polarization) of incident light, and the spatiotemporal phase between incident coherent pulses[26-30]. Therefore, demonstrating actively tunable directional excitation of highly squeezed polaritons, without resorting to the modulation of the source, remains an open challenge that is highly sought after due to its importance for the development of advanced nano-photonic technology.

Here we propose a new scheme, i.e., via controlling the sign of group velocity of polaritonic eigenmodes, to realize the tunable directional excitation of highly squeezed polaritons. When the optical source is fixed, the direction of excitation is then dictated by the sign of group velocity of the polaritonic eigenmodes. Hybrid plasmon-phonon polariton modes in graphene-$h$BN heterostructures[11-13] are chosen to exemplify the new scheme. This is because the strong coupling between plasmon and phonon polaritons in graphene-$h$BN heterostructures can flip the sign of group velocity of the resulting hybrid (plasmon-phonon polariton) modes[11-13] through tuning the chemical potential of graphene. Along with the advantage that the chemical potential of graphene can be tuned continuously via the electrostatic gating, graphene-$h$BN heterostructures is an attractive platform towards electrically reconfigurable nano-photonic circuits.

## 2. Results and discussion

### 2.1 Dispersion of hybrid plasmon-phonon polaritons in graphene-$h$BN heterostructures

In order to illustrate the underlying physics, we start with showing the dispersion of hybrid plasmon-phonon polaritons in graphene-$h$BN heterostructures in **Figure 1**. The $h$BN slab has a thickness of $d_{hBN} = 15$ nm and is encapsulated by two graphene layers (see Figure 1(a)); the thickness of $h$BN shall be judiciously chosen so that by tuning the chemical potential of graphene, we can control the sign of group velocity for the hybrid polaritons in a flexible way; see more discussion in **Figure S8**. The detailed model for graphene's surface conductivity and $h$BN's relative permittivity can be found in the supporting information. The surface optical conductivity of graphene is obtained from random phase approximation (RPA)[14]; an electron mobility of 10000 cm$^2$V$^{-1}$s$^{-1}$ is used to characterize the loss in graphene[14], where a smaller value of electron mobility represents a higher loss; the temperature in the RPA calculation is set to be 0 K; since the chosen working frequency (~24.2 THz) is much smaller than the corresponding frequency ($\geq 48$ THz) of the twice chemical potential of graphene (i.e., $2\mu_c \geq 0.2$ eV), the surface conductivities at 0 K and at 300 K will have small difference[14]. The experimental data with the consideration of realistic loss is adopted to model the uniaxial $h$BN[13-14], i.e., $\text{diag}(\epsilon_{r,x}, \epsilon_{r,y}, \epsilon_{r,z})$, where $\epsilon_{r,x} = \epsilon_{r,y}$ and $\epsilon_{r,z}$ are the components of relative permittivity parallel or perpendicular to the $h$BN plane (i.e. $x$-$y$ plane), respectively; since the reciprocal lattice vector of $h$BN is at least two orders of magnitude larger than the wavevector of polaritons in our discussed cases, it is reasonable to neglect the nonlocal effect of $h$BN[13,14].

The heterostructures is assumed to be on a dielectric substrate with relative permittivity $\epsilon_r = 2.4$ at the frequency of interest[31].

Figure 1(b,c) shows the dispersion of hybrid polaritons in graphene-hBN heterostructures under different chemical potentials of graphene in the first restrahlen band of hBN. Since hBN's phonon polaritons lie only within the reststrahlen band, the dispersion lines in Figure 1 existing outside the reststrahlen band is mainly due to the existence of graphene plasmons in the heterostructures. When the chemical potential of graphene is high, such as that in Figure 1(c), the hybrid polariton mode will be more like graphene plasmons [14]; see more detailed analysis in **Figure S2**. In addition, the *linear* shape of the dispersion line in Figure 1(c) is mainly due to the small frequency range (i.e., 22-25 THz) and the large wavevector range (i.e., 0-100 μm$^{-1}$) in the figure. The value of the group velocity $v_g = d\omega/dk$ for the hybrid polaritons in Figure 1(c) are within the range of $0.1c$-$0.3c$, where $c$ is the speed of light in free space. Regarding to the higher-order modes of the hybrid polaritons in Figure 1, they still exist in our designed heterostructures; see analysis in **Figure S3**.

For clarity, we define the sign of group velocity $v_g = d\omega/dk$ with respect to the phase velocity $v_p = \omega/k$: the sign of group velocity is positive (negative) when the group velocity and the phase velocity are in the same (opposite) direction. When the chemical potential of graphene is $\mu_c = 0.1$ eV in Figure 1(b), the hybrid polaritonic modes at the frequency of $\omega/2\pi = 24.2$ THz has a negative group velocity, and behaves more like the hBN's phonon polaritons[8-14]. On the contrary, when the chemical potential of graphene is increased to $\mu_c = 0.4$ eV in Figure 1(c), the group velocity of the hybrid polaritonic mode at 24.2 THz becomes positive, where the hybrid mode is more like graphene plasmons[3-5,11-14]; see Figure S2. We emphasize that the effect described here applies to hBN sandwiched by two graphene layers. For a monolayer graphene on hBN, the plasmon-phonon coupling in the first restrahlen band is much weaker[11,14]. This capability of controlling the sign of group velocity for these highly squeezed polaritons in a flexible way can enable the demonstration of many unique nanoscale applications. For example, such capability can enable the all-angle negative refraction between highly squeezed polaritons in graphene-hBN heterostructures[14]. Here we show that this capability can also lead to the directional excitation of highly squeezed polaritons in a controllable manner. Below the working frequency is set to be 24.2 THz according to Figure 1.

**2.2 Sommerfeld integration in polaritonic systems of graphene-hBN heterostructures**

In order to visualize the directional excitation of polaritons, the spatial distribution of fields in graphene-hBN heterostructures driven by a point source is analytically studied. For the directional excitation of polaritons in symmetric structures, a specific design of source is required, such as the widely exploited elliptically/circularly-polarized dipole[24-25,28-29]. Circularly-polarized dipoles can be practically realized through illuminating a nanostructure (such as a hole or a slit) with circularly polarized light[25,28]. In this work, a circularly polarized source with a dipole moment of $\bar{p} = \hat{x}p_x + \hat{z}p_z$ is introduced to excite the highly asymmetric polaritonic fields, where $p_x = i$ and $p_z = 1$; the corresponding spatial frequency spectra of the source in free space is shown in **Figure S4**. The point source is located at the original point and is above the hBN slab with a vertical distance of $d_0 = 1$ nm. According to the macroscopic electromagnetic theory, the field in each region in cylindrical coordinates $(\rho, \phi, z)$ is derived as follows:

$$H_{\phi m} = \int_0^\infty dk_\rho \left( \frac{-ip_z}{4\pi} \frac{k_\rho^2}{k_{z,1}} J_0'(k_\rho \rho) \pm \frac{p_x}{4\pi} \cos\phi \, k_\rho J_0''(k_\rho \rho) \right) A_m \left( e^{ik_{z,m}|z|} + \tilde{R}_{m,m+1} e^{ik_{z,m}z} \right) \qquad (1)$$

The coefficient $m$ in all subscripts is defined with the following rule: $m = 1$ when $z > -d_0$ (region of air), $m = 2$ when $-d_0 > z > -d_0 - d_{hBN}$ (region of hBN), and $m = 3$ when $z < -d_0 - d_{hBN}$ (region of substrate). The "+" ("-") sign in Equation 1 is adopted when $z > 0$ ($z < 0$). $k_\rho$ and $k_{z,m} = \sqrt{k_0^2 \frac{\epsilon_{r,x,m}}{\epsilon_{r,z,m}} - k_\rho^2}$ are the components of the wavevectors parallel and perpendicular to the interface. By enforcing the boundary conditions, the amplitude coefficients $A_m$ and the generalized reflection coefficients $\tilde{R}_{m,m+1}$ can be obtained analytically; see supporting information for details. $J_0'$ ($J_0''$) is the first (second) derivative of zero-order Bessel function of the first kind.

By performing the integration in Equation 1, the spatial distribution of fields excited by a circularly polarized dipole can be solved numerically. However, since there might be branch point singularities or polariton pole singularities at the real $k_\rho$-axis, the integration in Equation 1 is not well-defined. To overcome the ambiguity of integration from the branch point singularity, a small loss is usually assumed in the studied system (i.e., $\epsilon_{r,x,m}$ and $\epsilon_{r,z,m}$ are complex values). This way, the branch point will be lifted to the first quadrant of the complex $k_\rho$-plane. The integration path along the real $k_\rho$-axis near the branch point can then be safely detoured to the segment of ABCD in the fourth quadrant, as shown in **Figure 2**(a-b). This is exactly the scheme applied for the Sommerfeld integration in passive systems[14,28-29]. On the other hand, to overcome the ambiguity of integration from the polariton pole singularity[32-33], a similar procedure above can be adopted for the passive graphene-$h$BN heterostructures studied in this work. However, it should be emphasized that when the pole singularities correspond to the polariton eigenmodes having different signs of group velocities, different integration paths shall be judiciously chosen. When the sign of group velocity of polaritons is positive and a trivial loss is assumed, the polariton pole will show up in the first quadrant (instead of the real $k_\rho$ axis) of the complex $k_\rho$ plane. This way, we can safely detour the integration path along the real $k_\rho$-axis near the polariton pole to the segment of EF$_1$G$_1$H in the fourth quadrant, as shown in Figure 2(a). In contrast, when the sign of group velocity of polaritons is negative and a trivial loss appears, the polariton pole will emerge in the fourth quadrant of the complex $k_\rho$ plane. Then the integration along the real $k_\rho$-axis near the polariton pole shall be detoured to the segment of $EF_2G_2H$ in the first quadrant, as shown in Figure 2(b). Interestingly, with the judicious choice of the branch cut in Figure 2, the multi-value function of $k_{z,m}$ in Equation 1 is uniquely defined along the chosen integration path. Namely, the value of $\text{Im}(k_{z,m})$ along this newly chosen integration path is always positive. While $k_{z,m}$ is always in the first quadrant of the complex $k_{z,m}$ plane for the isotropic region of air or substrate; $k_{z,m}$ can be either in the first or second quadrant of the complex $k_{z,m}$ plane for the uniaxial region of $h$BN. Therefore, our detoured integration path is exactly the Sommerfeld integration path. By performing the Sommerfeld integration of Equation 1, the spatial distribution of highly squeezed polaritons in graphene-$h$BN heterostructure can be solved numerically.

**2.3. Directional excitation of hybrid polaritons in graphene-$h$BN heterostructures**

**Figure 3** shows the field distribution of tunable directional excitation of highly squeezed polaritons in graphene-$h$BN heterostructures. For clarity of conceptual demonstration, we begin by neglecting the loss in the $h$BN slab. When the chemical potential is low ($\mu_c = 0.1$ eV), most of the excited field flows to the left side of the source, while the part of the field flows to the right side of the source is negligible; see Figure 3(a). In contrast, when the chemical potential is high ($\mu_c = 0.4$ eV), most of the excited field flows to the right side of the source in Figure 3(b), different from Figure 3(a). From Figure 3, it is sufficient to use these two specific values of chemical potential of graphene to illustrate the phenomenon of tunable directional excitation of polaritons; for other values of chemical potential of graphene, the corresponding field distributions are shown in **Figures S6&S7**. The phenomena in Figure 3 can be explained from the perspective of the sign of group velocity as follows. When the source remained unchanged in Figure 3, the sign of the group velocity of the polaritonic eigenmodes has been flipped through changing the chemical potential from 0.1 eV in Figure 3(a) to 0.4 eV in Figure 3(b). For the circularly polarized source selected in Figure 3, it always tends to excite the polaritonic modes with the phase velocity towards the left side of the source (i.e., $\text{Re}(k_\rho) > 0$)[28]. However, the direction of energy flow is determined by the direction of the group velocity, instead of the phase velocity. Therefore, for the excited eigenmode, when its phase group and group velocity are in phase, most of energy will flow to the right side of the source, such as that in Figure 3(b); however, when its phase and group velocity are out of phase, most of energy will flow to the left side of the source, such as that in Figure 3(a). It shall be noted that while the directional excitation of polaritons has been extensively studied in symmetric structures with a positive group velocity of polaritons (similar to that in Figure 3(b)) through the control of polarization of the excitation source, the control of directionality of polaritons through flipping the sign of group velocity has not been explored. Since the chemical potential in graphene can be continuously tuned through electrostatic gating, the directional excitation of polaritons in graphene-hBN heterostructures can be controlled in an electrical way.

Due to the loss is inevitable in practical application, its influence on the new scheme of the group-velocity-controlled and gate-tunable directional excitation of polaritons is shown in Figure 3(c-d). While the appearance of loss in hBN will degrade the performance of the directional excitation, the proposed new scheme is robust to the realistic loss of hBN. In addition, the loss in hBN can be largely suppressed through isotopic enrichment of hBN[34]. This will reduce the influence of loss from hBN on the performance of the directional excitation. (The loss in graphene has been taken into consideration in Figure 3)

It should be noted that the graphene-hBN heterostructures or even a thin hBN slab alone can also enable the design of deep-subwavelength polaritonic dichroic splitter[20]; see supporting information. We use a thin *h*BN slab to exemplify the proposed dichroic splitter in **Figure S5**. When the working frequency is within the first (second) reststrahlen band of *h*BN, most of the excited energy will flow to the right (left) side of the source, see Figure S5(d) (Figure S5(e)). This is because the sign of group velocity of phonon polaritons are opposite in the first and second reststrahlen band of *h*BN, and the underlying mechanism in Figure S5 is the same as that in Figure 3. Therefore, *h*BN slabs or graphene-*h*BN heterostructures can be a promising platform to spatially separate the excited polaritons with different frequencies, with the latter offering the attractive option of electrical tunability.

**3. Conclusion**

In summary, we propose a new mechanism for the gate-tunable directional excitation of highly squeezed polaritons in graphene-*h*BN heterostructures, within the framework of classical electrodynamics. The tunable directional excitation of polaritons comes from the flexible control of the sign of group velocity of polaritonic eigenmodes, instead of the asymmetric structural design or the modulation of the incident source. Since the sign of group velocity of polaritons can be flexibly controlled by tuning the chemical potential of graphene through electrostatic gating, graphene-*h*BN heterostructures might be an attractive platform for the design of active nano-photonic devices and highly integrated systems with electrically reconfigurable advantages. Furthermore, it is noted that the tunable directional coupling has immediate applications in quantum optics. For example, in the chiral quantum optics, there is a great deal of interest in the directional control of emission by quantum emitters[35]. Since the quantum emitter is analogous to the point source (i.e., a dipole) considered in this work, the proposed scheme of gate-tunable directional coupling shall also be applicable within the framework of chiral quantum optics. This indicates graphene-

$h$BN heterostructure might also be a promising platform for the active quantum control of light-matter interaction and for the design of complex quantum networks.

**Supporting Information**

Additional supporting information may be found in the online version of this article at the publisher's website.

**Acknowledgements**

This work was sponsored by the National Natural Science Foundation of China under Grants No. 61625502, No. 61574127, No. 61601408, No. 61775193 and No. 11704332, the ZJNSF under Grant No. LY17F010008, the Top-Notch Young Talents Program of China, the Fundamental Research Funds for the Central Universities, and the Innovation Joint Research Center for Cyber-Physical-Society System, Nanyang Technological University under NAP Start-Up Grant, Nanyang Research Award (Young Investigator), and Singapore Ministry of Education under Grants No. Tier 1 RG174/16(S), No. MOE2015-T2-1-070 and No. MOE2016-T3-1-006, National Science Foundation under grant number NSF/EFRI-1741660.

**Keywords**

directional excitation, plasmon-phonon polaritons, graphene-$h$BN heterostructures

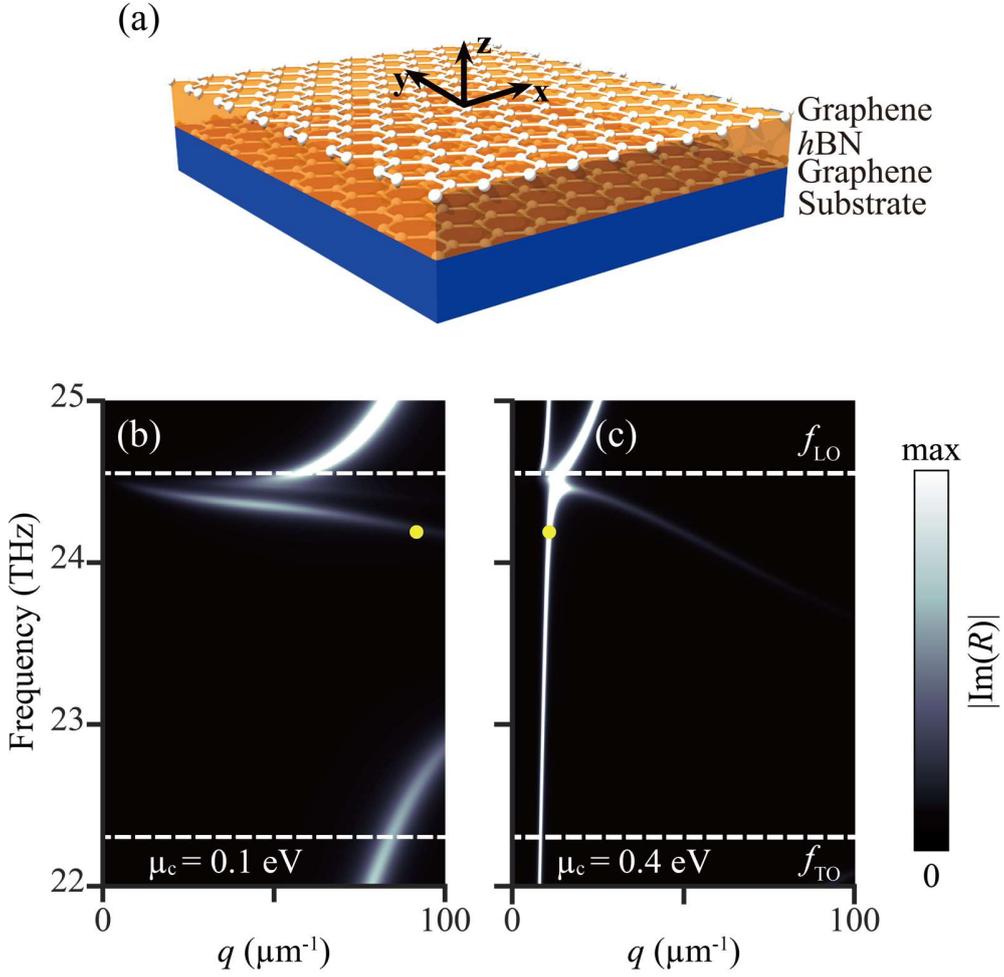

Figure 1. Hybrid plasmon-phonon polaritons in graphene-hBN heterostructures. (a) Structural schematic. A thin hBN slab with a thickness of 15 nm is encapsulated by two graphene layers. (b,c) Dispersion relation of hybrid plasmon-phonon polaritons. A false-color plot of $|\text{Im}(R)|$ is used to visualize the polaritonic dispersion, since the polaritons correspond to the singularity poles in $R$, where $R$ is the reflection coefficient of TM (or p-polarized) waves. For both graphene layers, the chemical potential is $\mu_c = 0.1$ eV in (b) and $\mu_c = 0.4$ eV in (c). The dashed lines correspond to the transverse optical (TO) and longitudinal optical (LO) frequencies, respectively. The group velocity of hybrid polaritons at 24.2 THz (highlight by yellow dots) is negative (positive) with respect to the phase velocity in (b) (in (c)).

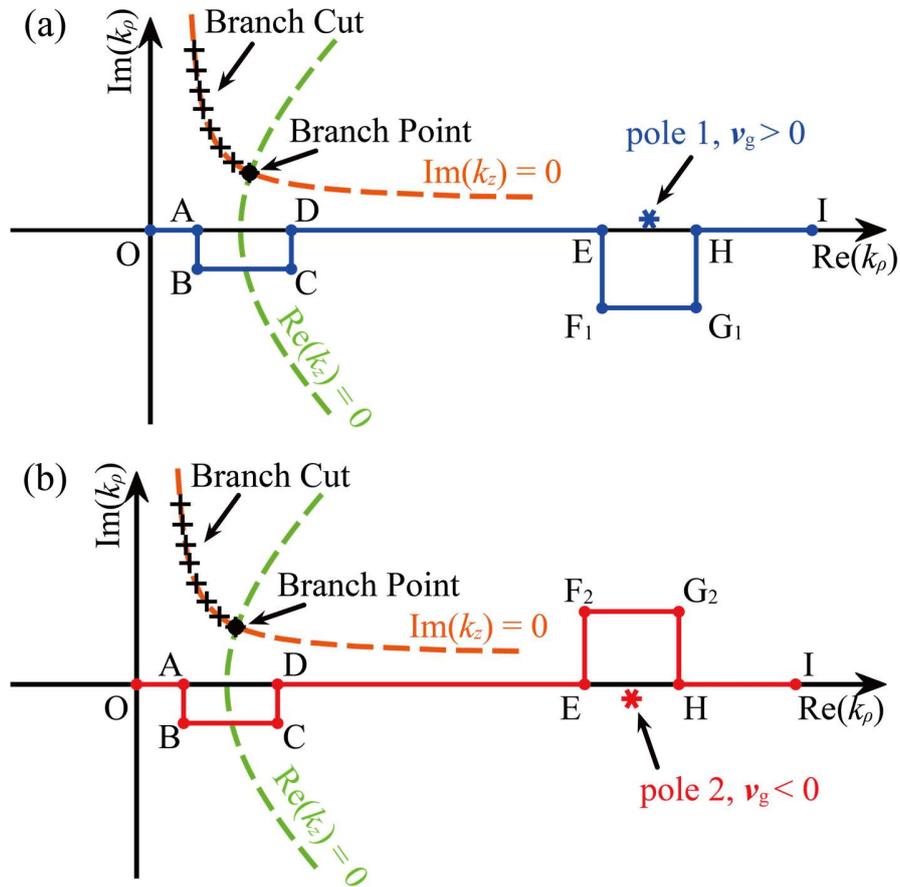

Figure 2. Choice of Sommerfeld integration path (blue and red lines) in systems with polaritons. When the group velocity of polaritons is (a) positive ((b) negative) with respect to the phase velocity, the pole singularity for the polaritonic mode is located at the first (fourth) quadrant of the complex $k_\rho$-plane. The Sommerfeld integration is considered in the real frequency domain.

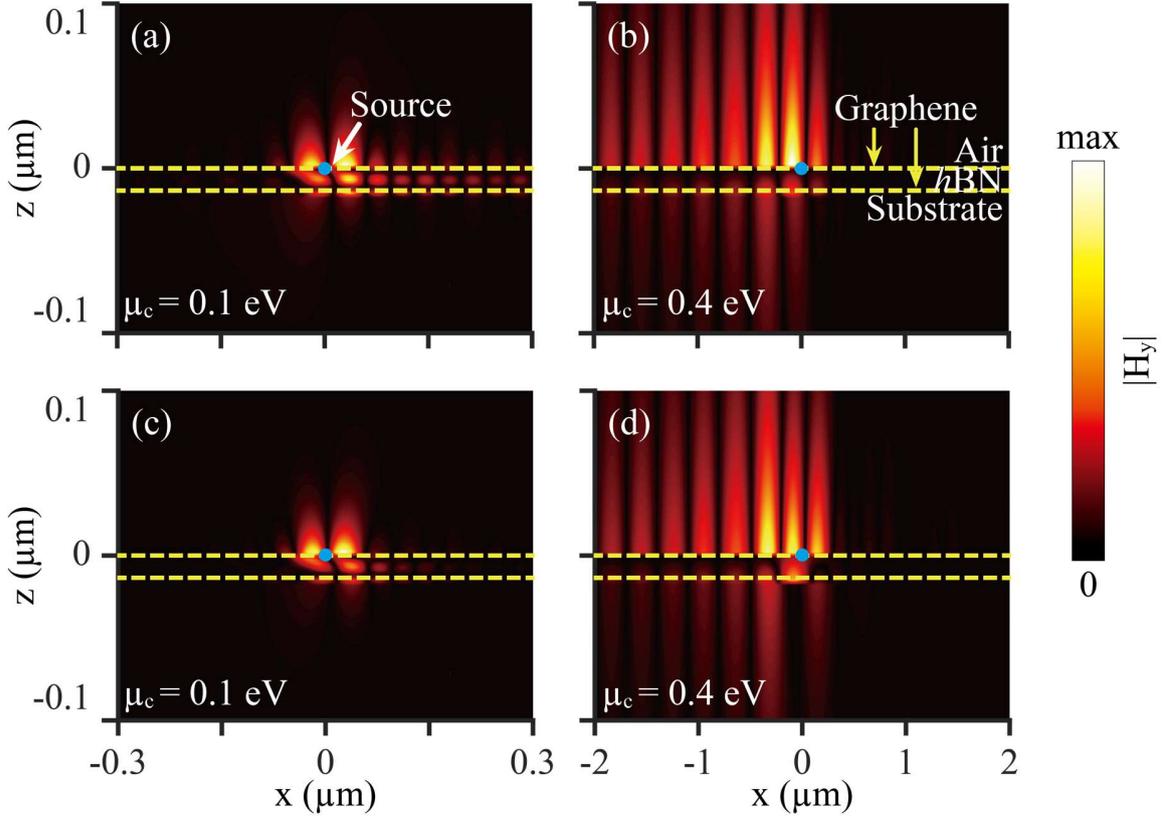

Figure 3. Field distribution of gate-tunable directional excitation of hybrid plasmon-phonon polaritons in graphene-hBN heterostructures at 24.2 THz. For conceptual demonstration, hBN is assumed to be lossless in (a, b), i.e., by artificially neglecting the imaginary part of hBN's permittivity. In (c,d), the realistic loss of hBN is considered. The source, is 1 nm above the hBN slab and has a dipole moment of $\bar{p} = [p_x, p_z]$, where $p_x = i$ and $p_z = 1$. The structural setup in (a,c) (in (b,d)) is the same as that in Figure 1(b) (in Figure 1(c)).